\documentclass[conference]{IEEEtran}
\IEEEoverridecommandlockouts
\usepackage{cite}
\usepackage{amsmath,amssymb,amsfonts}
\usepackage{algorithmic}
\usepackage{graphicx}
\usepackage{textcomp}
\usepackage{xcolor}
\usepackage{multirow}
\usepackage{caption}
\usepackage{subcaption}
\DeclareCaptionLabelFormat{cont}{#1~#2\alph{ContinuedFloat}}
\def\BibTeX{{\rm B\kern-.05em{\sc i\kern-.025em b}\kern-.08em
    T\kern-.1667em\lower.7ex\hbox{E}\kern-.125emX}}
\begin{document}

\title{A Compact Formulation for the Multi-UAV Assisted Data Gathering Problem in Wireless Sensor Networks : A Linear Programming Approach}

\author{\IEEEauthorblockN{Chi-Hieu Nguyen}
		\IEEEauthorblockA{Hanoi University of Science and Technology\\
			Ha Noi, Viet Nam \\
		hieu.nc151331@sis.hust.edu.vn}
\and
	    \IEEEauthorblockN{Khanh-Van Nguyen}
		\IEEEauthorblockA{Hanoi University of Science and Technology\\
			Ha Noi, Viet Nam \\
			vannk@soict.hust.edu.vn}
}

\maketitle

\begin{abstract}
In this paper, we study the problem of gathering data from large-scale wireless sensor networks using multiple unmanned air vehicles (UAVs) to gather data at designated  rendezvouses, where the goal is to maximize the network lifetime. Previous proposals often consider a practical approach where the problem of determining a data gathering scheme is decomposed into 2 sub-problems: i) partitioning the networks into clusters for determining the rendezvouses as these obtained cluster heads; and ii) determining the paths for a set of a given number of UAVs to come gathering data at these rendezvouses which have been harvesting data within each local clusters, respectively. We try to deal with this as a whole optimization problem, expecting a significant increase in computation complexity which would bring new challenge in creating practical solutions for large-scale WSNs. We introduce two alternatives mixed-integer linear programming (MILP) formulations, namely the 2-index model with $O(n^2)$ variables and the 3-index model that has $O(n^3)$ variables, where $n$ denotes the number of sensor nodes. We show that our best model could solve optimally the problem instances with up to 50 sensor nodes in less than 30 minutes. Next, we propose a heuristic idea to reduce the number of variables in implementing the 3-index model to effectively handle larger-scale networks with size in hundreds. The experiment results show that our heuristic approach significantly prolongs the network lifetime compared to existing most efficient proposals.
\end{abstract}

\section{Introduction}
Wireless sensor networks (WSNs) play an important role in the Internet of Things (IoT) for implementing real-time monitoring or long-term surveillance purposes. In large-scale WSN, several thousands of sensor nodes powered by batteries are deployed in a vast region and they coordinate to perform two main tasks: sensing and communicating. Traditionally, the sensory data must be periodically sent to a static sink node through multi-hops communication. However, this approach is suffering from the hot spot problem where the energy of nodes nearer to the sink depletes faster due to the huge traffic load from far away nodes ~\cite{7004042}. To tackle this issue, employing one or more mobile data collectors as the mobile sinks to move around and gather sensory data has been considered as an efficient approach to reduce and also balance the energy consumption between SNs thus prolong the network lifetime ~\cite{6359890}.

Recently, the Unmanned Aerial Vehicles (UAVs) technology have seen rapid growth with application in various fields such as military, emergency rescue, transportation, surveillance and monitoring. This technology has also emerged as a flexible and convenient approach to collect the harvested info  data in large-scale WSNs because of their advantages such as low cost, simple deployment and ability to avoid terrain obstacles compared with other types of mobile collectors. In additional, the ground-to-air communication links between sensor node (SN) and UAVs typically exploit the line-of-sight (LoS) propagation thus the data transmission is more reliable and also provide higher data rate compared to ground vehicles.
However, the critical issue in this technology is that the UAVs usually have limited endurance due to the practical physical constraints, and need to be collected for battery swap or recharging, e.g., the endurance time can be 30 minutes for a typical rotary-wing UAV~\cite{UAVendurJSAC18}.

Similar to other kinds of mobile collector, recent research works \cite{ZZZ18WCL,RPPiWSN08TrsMoC,Dong2014,Alfattani19GloCom,WU19ComNet} in UAV-assisted data gathering can be classified into two basic approaches: rendezvous-less and rendezvous-based. The former approach requires the UAVs to visit every SN neighborhood at least once so that each SN can send its data directly to a UAV via single-hop transmission. Although the rendezvous-less approach improves the communication reliability and energy conservation of SNs, the length of UAV trajectory becomes excessively long thus this approach is often infeasible for large-scale WSNs. This challenging issue can be resolved by using the rendezvous-based approach, where the UAVs only visit a subset of predefined nodes called rendezvous nodes to gather data from their nearby SNs. 

Several works \cite{KASWAN17, Ebrahimi18, Alfattani19GloCom, KONSTANTOPOULOS18ComNet} on the rendezvous-based data gathering have been proposed and most of them follow a common idea of decomposing the main problem into two sub-problems which will be solved one by one: i) designing a clustering scheme wherein the cluster heads (CHs) take the responsibility as rendezvous nodes to gather data from each cluster; ii) determining UAV trajectories for collecting the data at the CHs and transporting to the base sink. Although these sub-problems are easier to deal with, this method can affect the solution quality.

 In this paper, we address at the same time the optimization problem of both the network clustering and UAV trajectory designing for a data collection period (or a round in other words) in WSN. Our objective is to maximize the network lifetime, which can be employed in each round by using an alternative objective function of SN energy. To this end, we first give a formulation of the problem under a mixed-integer linear programming model. Although there exist previous works in the literature that study the UAV-assisted data gathering problem using linear programming approach, however, they are not general enough that the proposed model considers only the scenario of a single UAV as in \cite{Yang2008, Zhao2012, Zhan2017, ANWIT2020106802} or follow the rendezvous-less approach as in \cite{s17040742}. To the best of our knowledge, this work is the first attempting to propose a MILP formulation for the rendezvous-based data gathering problem with multiple UAVs in WSN. Our contributions are as follows:

\begin{itemize}
    \item We introduce two formulations of the problem under a mixed-integer linear programming model, namely the 2-index and 3-index models. Then we show that the 3-index model can bring significant savings in computational effort compared to the 2-index model, even though it has a higher number of variables and constraints.
    \item We propose a heuristic model that reduces the number of variables in the 3-index model to solve the problem instance representing a practical larger-scale network with size in hundreds.
    \item We conduct experiments to demonstrate our heuristic model's effectiveness in terms of extending network lifetime compared to the existing proposal.
\end{itemize}

The rest of this paper is organized as follows. Section II describes the system model. Section III presents two proposed MILP formulations for the problem. Section IV discusses our numerical and experimental results. Section V concludes the paper.

\section{System model}

We consider a sensor network consisting of a sink node $s_0$ and a set of $n$ sensor nodes (SNs) $\mathcal{S} = \{s_1,...,s_n\}$. Assume that the data of all sensors must be collected at the sink periodically with a completion time required of $T_{max}$ seconds, which is determined from the application requirements. Each period of data collection is called a round and $T_{max}$ is called the round length. There are $m$ rotate wing UAVs deployed to collect data from sensors, all fly at a fixed altitude $h$. In each round, the network is partitioned into clusters and each SN transmits its data to an associated CH in a multi-hop fashion. Then each UAV departs from the sink, sequentially visits some specified CHs to collect data and finally return to the sink. Assume that the residual energy of SN $s_i$ after the $r^{th}$ round is $E_i^r$ the amount of data to collect at $s_i$ is $b_i$. The problem asks for a data collection scheme in the $(r+1)^{th}$ round including $k \geq 1$ cluster trees and $m$ UAV trajectories so that the minimum SN energy after $(r+1)^{th}$ round is maximized, while satisfying the following:

\begin{itemize}
    \item Each cluster tree is rooted at the CH and spans all cluster nodes.
    \item Each SN is included in exactly one cluster tree.
    \item Each UAV trajectory is a ring that contains the sink.
    \item Each CH is included in exactly one UAV trajectory.
    \item Each UAV trajectory has length less than a given value $L_{max}$
\end{itemize}

We utilize the same radio model as discussed in~\cite{926982}. In this model, the free space ($d^2$ power loss) model is used if the distance between the transmitter and the receiver is less than a $d_0$ threshold, otherwise the model of multi-path ($d^4$ power loss) is used. Thus, the energy spent to transmit a $l$-bits packet over distance $d$ is:
\begin{equation}
E_{Tx}(l,d) = 
\begin{cases}
    lE_{elec} + l\epsilon_{fs}d^2,& d < d_0\\
    lE_{elec} + l\epsilon_{mp}d^4,& d \geq d_0
\end{cases}
\end{equation}
The energy spent by the radio to receive $l$-bits of data is:
\begin{equation}
E_{Rx}(l) = lE_{elec}
\end{equation}
For more convenient, let $E_{Rx} = E_{elec}$ be the per-bit receiving energy of a SN and $E_{Tx}^{ij} = E_{elec} + \epsilon_{fs}d_{ij}^2$ be the per-bit transmitting energy from node $s_i$ to $s_j$, respectively. Thus, if node $s_i$ has to relay $B_i$ bits of data from the descendants in its cluster tree in a round, then the total energy consumption of $s_i$ after the round is calculated as:
%Moreover, we assume that . 
\begin{equation}
\begin{split}
E_i^c & = E_{Rx}(B_i) + E_{Tx}(B_i + b_i) \\
    &  = (E_{Rx} + E_{Tx}^{ij}) B_i + E_{Tx}^{ij} b_i
\end{split}
\end{equation}
The remaining energy of $s_i$ after the $(r+1)^{th}$ round is
\begin{equation}
E_i^{r+1} = E_i^r - E_i^c
\end{equation}

The multi-UAVs data collection problem asks for $m$ data forwarding trees corresponding to $m$ clusters, each of which roots at a CH node and $k$ UAV trajectories which start and end at the sink. The objective is to maximize the network lifetime, which is defined as the time span from deployment to the
instant when the first SN run out of energy.

\section{MILP Formulations}

\subsection{Objective function and basic constraints}
Both formulations proposed in this section used the same objective function and shared some basic variables and constraints. We will first present these fundamentals in this subsection.

In both formulations, the cluster configuration (i.e the network spanning forest) and all UAV trajectories will be defined by two types of binary variables $x_{ij}$ and $r_{ij}$. In particular, the variable $x_{ij}$ takes value 1 when $s_i$ and $s_j$ are in the same cluster and $s_j$ is the parent node of $s_i$ in the data forwarding tree, otherwise $x_{ij}$ is set to 0. Next, the binary variable $r_i$ takes value 1 if $s_i$ and $s_j$ are two CHs visited by the same UAV and $s_j$ is visited right after $s_i$ on the trajectory, or $r_{ij} = 0$ if one of these above conditions is violated.

Since we aim to maximize the network lifetime, the objective function in the MILP model must be relevant to the residual energy of SNs. Thus, we define $n+1$ real variables $e_1, e_2,..., e_n, e_{min}$ where the first $n$ variables equal to $n$ values of SN's residual energy, respectively and the $(n+1)^{th}$ variables that will take the minimum value among these former.

With these above variables, the problem objective and basic constraints are given from (1) to (5).

$$
\begin{array}{l@{}lr}
\textbf{Objective:$\;\;\;\;\;\;$} & \displaystyle \max f = \alpha e_{min} + (1-\alpha)\frac{1}{N} \sum_{i \in V^*} e_i & \text{(O)}\\
\textbf{Constraints:} &  & \\
\end{array}
$$

$$
\begin{array}{clr}
\displaystyle \sum_{j \in V \setminus \{i\}} x_{ij} + \sum_{j \in V \setminus \{i\}} r_{ij} = 1 & \forall i \in V^* & \text{(B.1)}\\
\displaystyle \sum_{j \in V \setminus \{i\}} r_{ij} = \sum_{j \in V \setminus \{i\}} r_{ji} & \forall i \in V^* & \text{(B.2)}\\
\displaystyle \sum_{j \in V^*} r_{0j} \leq m & & \text{(B.3)}\\
x_{ij} = 0 & \forall i,j \in V & \\
&\text{and } d_{ij} > R & \text{(B.4)}\\
x_{ij}, r_{ij} \in \{0,1\} & \forall i,j \in V & \text{(B.5)}\\
e_{min} \geq 0, e_i \geq e_{min} & \forall i \in V & \text{(B.6)}\\
\end{array}
$$

As shown in (O), the objective of the MILP model is to maximize a weighted sum of $e_{min}$ and $\frac{1}{N} \sum_{i \in V^*} e_i$, that are respectively the minimum and the average residual energy of SNs. Intuitively, since the network lifetime is equal to the time until the value of $e_{min}$ reaches 0, we want to decrease the reduction rate of $e_{min}$ in the long-term thus our approach is to greedily maximize the value of $e_{min}$ after each round. However, we also need to consider the average residual energy in the network to avoid the case of inefficient consumption at other SN with higher energy. 

\begin{figure}[htbp]
    %\captionsetup{belowskip=0pt}
	%\setlength{\belowcaptionskip}{-10pt}
    \centerline{\includegraphics[width=3 in]{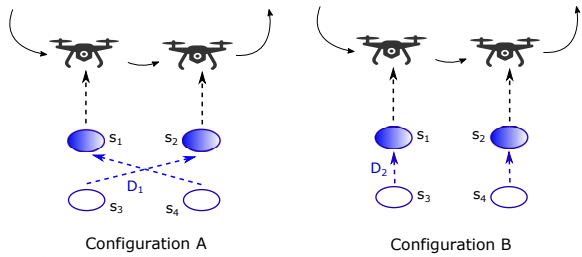}}
    \caption{An example of two cluster configurations.}
    \label{config_example}
\end{figure}

For example, Fig.~\ref{config_example} illustrates two different cluster configuration for a network scenario where $s_1$ and $s_2$ are CHs. In this example, we assume that all SNs start with the same energy and generate the same 1 byte of data. Both cluster configurations yield the same value of $e_{min}$ which equal the residual energy of the CH nodes since each of them consume the same amount of energy to receive 1 byte and transmit 2 bytes to the UAV. However, the second configuration is more efficient because the other nodes $s_3$ and $s_4$ consumed less energy due to shorter transmission range. Therefore, by including the average SN energy in the objective as in (O), the better configuration is chosen. 

In section \ref{sec:experiment}, we will do the experiment to find the optimum weight value $\alpha$ so that the network lifetime is maximized.

Next, consider the basic constraints, constraint (B.1) means that a SN must be either forward data to its parent node if it is not a CH ($\displaystyle \sum_{j \in V \setminus \{i\}} x_{ij} = 1$) or else forward data to a corresponding UAV ($\displaystyle \sum_{j \in V \setminus \{i\}} r_{ij} = 1$). Constraint (B.2) require that number of UAVs that arrive and leave a SN must be the same, namely equal to 1 if it is a CH and equal to 0 otherwise. The maximum number of UAVs is met due to (B.3). Finally, constraint (B.4) indicates that a SN $s_i$ could send data to $s_j$ only if the Euclide distance $d_{ij}$ between them is not exceed the maximum radio range $R$.

In the next two subsection, we present two completed MILP formulations for the multi-UAV data collection problem by adding other constraints on SN energy and trajectory length of UAV.

\subsection{2-index formulation}

In this formulation, we define the integer variables $B^n_i$ and $B^e_{ij}$ to calculate the energy expense on each SN. In particular, $B^n_i$ count the amount of data in bits that $s_i$ received from its descendants, while $B^e_{ij}$ count the amount of data sent from $s_i$ to $s_j$. The constraints related to $B^n_i$ and $B^e_{ij}$ are given in (F1.1) and (F1.2).

$$
\begin{array}{clr}
\displaystyle \sum_{j \in V} B^e_{ji} = B^n_i & \forall i \in V^* &
\text{(F1.1)}\\
B^n_i + b_i \leq B^e_{ij} + M(1 - x_{ij}) & \forall i \in V^* & \\
& ,j \in V & \text{(F1.2)}\\
\end{array}
$$

Constraint (F1.1) means that the amount of data that node $s_i$ received must be equal to the total amount of data sent through all of its incoming link. In (F1.2), the big-M notation is used to formulate a disjunctive constraint. Specifically, if $s_j$ is the parent node of $s_i$ ($x_{ij} = 1$) then the amount of data sent through the link $s_i - s_j$ must not be less than the total amount that $s_i$ received $B^n_i$ plus the amount generated by itself $b_i$. Otherwise, if $x_{ij} = 0$ then the inequality (8) becomes $B^n_i + b_i \leq B^e_{ij} + M$ and always gets true since $M$ is chosen as a very large value.

By applying the results in (3) with the variables $B^n_i$ and $B^e_{ij}$, the energy consumption at SNs can be formulated as follow:

$$
\begin{array}{cl}
\displaystyle E^r_i - E_{Rx} B^n_i - \sum_{j \in V^* \setminus \{i\}} E_{Tx}^{ij} B^e_{ij} & \geq e_i\\
& \forall i \in V^* \;\;\;\; \text{(F1.3)}\\
\displaystyle E^r_i - E_{Rx} B^n_i - E_{Tx}^{UAV} (B^n_i+b_i) & \\
\;\;\;\;\;\;\;\;\;\;\;\;\;\;\;\;\; \displaystyle \geq e_i - M\left(1 - \sum_{j \in V \setminus \{i\}} r_{ij}\right) & \forall i \in V^* \;\;\;\; \text{(F1.4)}\\
\end{array}
$$

Here in (F1.3), the residual energy of a SN $s_i$ must be less or equal than the initial value subtracted to the receiving energy ($E_{Rx} B^n_i$) and the total transmitting energy on the outgoing links ($\sum_{j \in V^* \setminus \{i\}} E_{Tx}^{ij} B^e_{ij}$). The constraint (F1.3) can be applied for all SNs, however it only takes into account the energy consumption for the radio communication between SNs. Then for the CHs, we also need to consider the energy spent to transmit data to the UAV. This is done by constraint (F1.4), which is only applied with the CH node where $\sum_{j \in V \setminus \{i\}} r_{ij} = 1$ by using big-M formulation.

Consider the constraint on UAV trajectory, we introduce the real variables $L_i$ corresponding to each SN $s_i$. If $s_i$ is CH then $L_i$ is equal to the length that its corresponding UAV (i.e. the UAV assigned with $s_i$) has flown from departure until it arrive at the hovering position above $s_i$. In the opposite case, the value of $L_i$ is set to 0. Therefore, if $s_i$ and $s_j$ are two consecutive nodes in a UAV trajectory (i.e. $r_{ij} = 1$), the value of $L_j$ is equal to $L_i$ plus the Euclide distance $d_{ij}$ between these two nodes. As a consequence, the total length of an UAV trajectory is calculated as $L_i + d_{i0}$ where $s_i$ is the last node in the trajectory (this condition is equivalent to $r_{i0} = 1$). Finally, the linear constraints related to the UAV trajectory are given in (F1.5) and (F1.6).

$$
\begin{array}{clr}
L_i + d_{ij} \leq L_j + M(1 - r_{ij}) & \forall i,j \in V & \text{(F1.5)}\\
\displaystyle L_i + d_{i0}  r_{i0} \leq L_{max} & \forall i \in V & \text{(F1.6)}\\
\end{array}
$$

In summary, the 2-index formulation for the data collection problem with multiple UAVs is:

$$
\begin{array}{l@{}lr}
\textbf{Objective$\;\;\;\;\;\;$} & \displaystyle \max f & \text{(F1)}\\
\textbf{subject to } & \text{(B.1) - (B.6) and (F1.1) - (F1.6)} & \\
\end{array}
$$

%Constraint (P1.2) means that an 

%Constraint indicates that a Sm must forward its data for once time to the parent node if it is not a ch or a nav otherwise

%Constraint require that number of incoming and outgoing uav at a sn must be equal. Namely, the value is equal to 0 if s -; is not a ch, other wise the value is 1 (in coder t on the saint)

%Constraint ensures that a sn could send data to another one if and only if they are in the communication radius r of each other

%The directed ring tree property of in-degree one or at most one for non-distributor nodes is enforced by constraints

%Intuitively, we can define the objective function is to maximize the minimum since our objective is maximize the network lifetime, 

\subsection{3-index formulation}
Although the above formulation is compact which only has $O(n^2)$ variables and constraints, however it is based on the big-M notation. It is known that the big-M contraint typically yields a very poor lower bound in the search trees and thus leading to a slower solving time ~\cite{}. Thus, in this subsection, we present a novel MILP formulation that can exclude the big-M notations.

This formulation is inspired by the multi-commodity flow formulation for the Ring-Tree problem in~\cite{Hill2018}. Here we assume that there are $n$ flows of commodity, each of them originates from a SN and terminates at the sink node. We call the flow that start from a source node $s_k$ as the $k^{th}$ commodity flow. Then the continuous flow variable $f_{ij}^k \in [0,1]$ is introduced to model the amount of the $k^{th}$ flow that is transferred from $s_i$ to $s_j$. All flow variables must be satisfied the following constraints:

$$
\begin{array}{clr}
f_{ij}^k \leq x_{ij} + r_{ij} & \forall i,j,k \in V & \text{(F2.1)}\\
\displaystyle \sum_{j \in V^*} f_{0j}^k - \sum_{j \in V^*} f_{j0}^k = -1 & \forall k \in V^* & \text{(F2.2)}\\
\displaystyle \sum_{j \in V \setminus \{k\}} f_{kj}^k - \sum_{j \in V \setminus \{k\}} f_{jk}^k = 1 & \forall k \in V^* & \text{(F2.3)}\\
\displaystyle \sum_{j \in V \setminus \{i\}} f_{ij}^k - \sum_{j \in V \setminus \{i\}} f_{ji}^k = 0 & \forall i,k \in V^* & \text{(F2.4)}\\
f_{ij}^k \in [0,1] & \forall i,j,k \in V & \text{(F2.5)}\\
\end{array}
$$

Constraint (F2.1) means that a commodity flow can only be transferred from $s_i$ to $s_j$ if the edge $(s_i,s_j)$ is either in a cluster tree ($x_{ij} = 1$) or in a UAV trajectory ($r_{ij} = 1$). Constraints (F2.2), (F2.3) and (F2.4) are the flow conservation constraints at the sink node, the source node and the transit nodes, respectively.

The flow-based model helps to eliminate the case where a solution contains subtour in a cluster or UAV trajectory since every flows must be end at the sink node. Moreover, these flow variables could be used to formulate other specific constraints of our problem, namely we can calculate the energy consumption of a SN based on the amount of flow that go in and out at that node and calculate the UAV trajectory lengths based on the paths of some certain flows.

First, consider the energy constraints, we need to calculate the amount of data that a SN received and sent. This is directly related to measuring the amount of flow that go in and out at a SN. However, we should notice that there are two types of flow at a SN $s_i$: the intra-cluster flow that originated from a SN $s_j$ in the same cluster with $s_i$ and the inter-cluster flow that comes from a different cluster. In the real network context, an intra-cluster flow indicates that there is a data transmission between SNs, while an inter-cluster flow can be seen as an UAV movement between two CHs. 

\begin{figure}[htbp]
    \captionsetup{belowskip=0pt}
	\setlength{\belowcaptionskip}{-10pt}
    \centerline{\includegraphics[width=2 in]{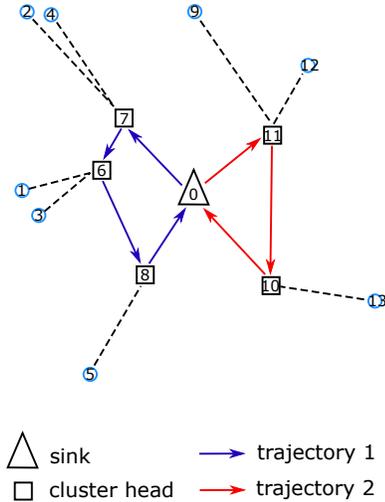}}
    \caption{Example of intra-cluster and inter-cluster flow.}
    \label{flow_example}
\end{figure}

Accordingly, we define the second flow variables $g_{ij}^k$ to measure the intra-cluster flows transferred between nodes. It is easy to see that a flow transferred from $s_i$ to $s_j$ is an inter-cluster flow if and only if $s_i$ and $s_j$ are two consecutive nodes on a UAV trajectory or when $r_{ij} = 1$ in other word (as illustrated in Fig.~\ref{flow_example}). Thus, the mathematical form of $g_{ij}^k$ can be written as:

\begin{equation}
g_{ij}^k = 
\begin{cases}
    f_{ij}^k & \text{if } r_{ij} = 0\\
    0 & \text{if } r_{ij} = 1
\end{cases}
\label{eqn:intra}
\end{equation}

The above formula can be translated to the linear constraints as follow:

$$
\begin{array}{clr}
g_{ij}^k \leq f_{ij}^k & \forall i,j,k \in V & \text{(F2.6)}\\
g_{ij}^k \geq f_{ij}^k - r_{ij} & \forall i,j,k \in V & \text{(F2.7)}\\
g_{ij}^k \leq 1 - r_{ij} & \forall i,j,k \in V & \text{(F2.8)}\\
g_{ij}^k \geq 0 & \forall i,j,k \in V & \text{(F2.9)}\\
\end{array}
$$

More details, when $r_{ij} = 0$ then the inequality (F2.7) become $g_{ij}^k \geq f_{ij}^k$ and by combining with (F2.6) we have $g_{ij}^k = f_{ij}^k$. Similarly, when $r_{ij} = 0$ then we have $g_{ij}^k \leq 0$ from (F2.8) and $g_{ij}^k \geq 0$ from (F2.9), thus $g_{ij}^k = 0$. In both case, the value of $g_{ij}^k$ always satisfies (\ref{eqn:intra}).

With the introduction of intra-cluster flow variables $g_{ij}^k$, the constraints on energy consumption of SNs are:

$$
\begin{array}{cl}
\displaystyle E^r_i - E_{Rx} \sum_{j \in V \setminus \{i\}} \sum_{k \in V^*} b_k g_{ji}^k & \\
\displaystyle \;\;\;\;\;\;\;\; - \sum_{j \in V \setminus \{i\}} E_{Tx}^{ij} \left( \sum_{k \in V^*} b_k g_{ij}^k \right) \geq e_i & \forall i \in V^* \;\;\; \text{(F2.10)}\\
\end{array}
$$

$$
\begin{array}{cl}
\displaystyle E^r_i - (E_{Rx} + E_{Tx}^{UAV}) \sum_{j \in V \setminus \{i\}} \sum_{k \in V^* \setminus \{i\}} & b_k (g_{ji}^k - g_{ij}^k) \\
\displaystyle  - E_{Tx}^{UAV} b_i \sum_{j \in V \setminus \{i\}} r_{ij} \geq e_i & \forall i \in V^* \;\;\; \text{(F2.11)}\\
\end{array}
$$

%- \sum_{j \in V \setminus \{i\}} E_{Tx}^{ij} \left(\sum_{k \in V^*} b_k g_{ij}^k\right)

Next, the constraints on UAV trajectory lengths are given in (F2.12)

$$
\begin{array}{c}
\displaystyle d_{0k} \sum_{j \in V \setminus \{i\}} r_{ij} + d_{ij} \sum_{j \in V \setminus \{i\}} (f_{ij}^k - g_{ij}^k) \leq L_{max}
\end{array}
$$
$$
\begin{array}{r}
\;\;\;\;\;\;\;\;\;\;\;\;\;\;\;\;\;\;\;\;\;\;\;\;\;\;\;\;\;\;\;\;\;\;\;\;\;\;\;\;\;\;\;\; \forall k \in V^* \;\;\; \text{(F2.12)}
\end{array}
$$

To sum up, the proposed 3-index formulation is as follow:

$$
\begin{array}{l@{}lr}
\textbf{Objective$\;\;\;\;\;\;$} & \displaystyle \max f & \text{(F2)}\\
\textbf{subject to } & \text{(B.1) - (B.6) and (F2.1) - (F2.12)} & \\
\end{array}
$$

\subsection{Formulation analysis}
Both presented formulation employ $2n^2+n+1$ base variables including $n^2$ variables $x_{ij}$, $n^2$ variables $r_{ij}$ and $n+1$ variables $e_{min}, e_1,..., e_n$. In the 2-index formulation, there are an additional of $n^2+2n$ variables including $Bn_i, Be_{ij}$ and $L_i$. Thus there are in total $3n^2+3n+1$ variables in this formulation. For the 3-index formulation, we introduce $2n^3$ flow variables $f^k_{ij}$ and $g^k_{ij}$. However, since we have $x_{ij} = 0$ and $g^k_{ij} = 0$ for all $i,j$ such that $d_{ij} > R$ (i.e. $s_j$ is out of transmission range of $s_i$) then we can reformulate the 3-index formulation so that there are only $n\delta$ variables $x_{ij}$ and $n^2\delta$ variables $g^k_{ij}$ where $\delta$ is the maximum number of neighbor nodes of a SN. In general, the 3-index formulation has higher number of $O(n^3)$ variables and constraints compared to $O(n^2)$ in the 2-index formulation, however we will show in the experiment that the solving time of the 3-index formulation is significantly faster than the other one because it can exclude the big-$M$ notation.

\subsection{Heuristic model}
Through the experiment, we find that our best MILP model can only deal with the problem of moderate size, namely up to 50 SNs thus it is insufficient to apply in a large-scale network of several hundred or thoudsands of nodes. Thus we propose a simple heuristic to reduce the number of variables in the MILP model. Since our objective is to prolong the network lifetime, the selected CHs in a round should be the nodes with higher residual energy. Therefore, we restrict the set of potential nodes to become CHs to the set of $P$ percent of nodes with the highest energy in the network $N_P$. As a result, the number of variables $r_{ij}$ can be reduced from $n^2$ to $|N_P|^2$ because $r_{ij} = 0$ if either i or j is not in the set of potential CHs $N_P$ and the number of variables $f^k_{ij}$ can also be reduced from $n^3$ to $|N_P|^3$. 

The first competitor is the Convex-Hull based protocol (CHP) in [10], which also handled the case of multiple UAVs. This protocol firstly divide the network into multiple sectors with the same percentage of energy decrease and each mobile collector (MC) is associated with an unique sector. An UAV trajectory is constructed inside its sector at initial by a geometrical method and then a simulated-annealing adjustment is adopted to improve the trajectories in term of total network energy consumption.
\section{Experiments}
\label{sec:experiment}
In this section, we first compare internally between two proposed MILP models and the heuristic implementation regarding the computation time and optimality. Later we evaluate the network lifetime performance when applying the heuristic model in comparison with an existing work. In the experiments, the network size is varied from 15 to 100 nodes. The communication range of a sensor node is 40m, and the flying altitude $H$ of each UAV is 20m. By default, each SN $s_i$ generates a data packet of a fixed size of 100Kb per round. Each SN's initial energy is randomly selected from the interval [20J, 40J], thus reflecting heterogeneous WSNs. Table~\ref{t_settings} sums up the experiment settings and parameters. 

All MILP models were solved by using the commercial MIP solver CPLEX in version 20.10. The experiments were run on a computer with an Intel(R) Core(TM) i7-4770 CPU @ 3.40GHz and 8 GB of RAM. For each test instance, the relative MIP gap tolerance is set to 0.1\%. We also set a solving time limit of 30 minutes in CPU time. Then if the running time for a test instance exceeds the predefined time limit, we terminate the solving process and record the best gap achieved. Other configuration parameters of the CPLEX solver are left as default.

\begin{table}[h]
    \centering\label{t_settings}
    \caption{Default experiment settings}
    \begin{tabular}{| l | l | l |}
        \hline
\multirow{5}*{Network} & \multirow{2}*{Area} & 15/20/30 SNs: 50m $\times$ 50m  \\
&  & 50 SNs: 75m $\times$ 75m  \\
&  & 100 SNs: 100m $\times$ 100m  \\ \cline{2-3}
& Sink location & Center of the square area   \\ \cline{2-3}
& Number of SNs & 15 - 100   \\ \cline{2-3}
& Number of UAVs & 2   \\ \cline{2-3}
& Max. trajectory length & 100m   \\ \cline{2-3}
& Packet size & 100Kb   \\ \cline{2-3}
& SN energy & Random in [20J, 40J]   \\ \hline
\multirow{3}*{Radio} & Communication range & 40m \\ \cline{2-3}
& $E_{elec}$ & 50 nJ/bit \\ \cline{2-3}
& $\epsilon_{fs}$ & 10 pJ/bit/m$^2$ \\ \cline{2-3}
& $\epsilon_{mp}$ & 0.0013 pJ/bit/m$^4$ \\ \hline
    \end{tabular}
    \label{t_settings}
\end{table}
\vspace*{-5pt}

\subsection{Impact of the weight parameter $\alpha$}
\begin{figure}[t!]
	\centering
	\setlength{\belowcaptionskip}{-10pt}
	\includegraphics[width=0.5\textwidth]{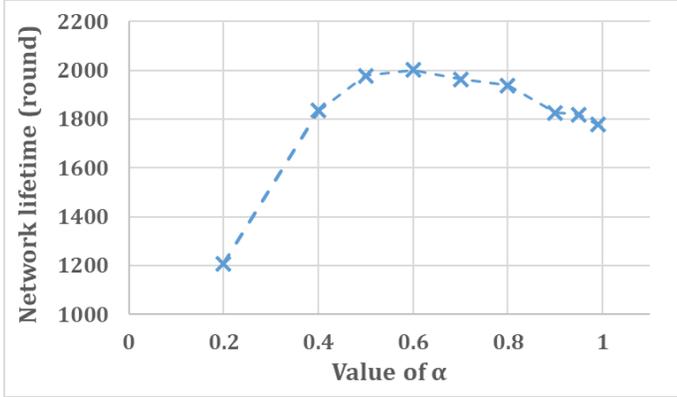}
	\caption{Impact of the weight $\alpha$ to the network lifetime}
	\label{experiment::alpha}
\end{figure}

Firstly, we examine the most efficient value of the weight factor $\alpha$ in the objective function that would induce the longest network lifetime. In particular, we experiment on a random WSN with 30 SNs and the value $\alpha$ is varied from 0.1 to 1. The network keep operating until the first SN is exhausted and we apply the 2-index model to obtain the network configuration in each round. Besides, we use a smaller data packet size of 10Kb for all SNs to differentiate between results more clearly. Fig~\ref{experiment::alpha} depicts the experiment results of network lifetime with different value of $\alpha$. As can be seen, the value $\alpha$ equal to 0.6 achieved the longest lifetime, namely approximate 2000 rounds. Accordingly, we will used the value $\alpha = 0.6$ in the next experiments.
%15 over 20 test instances. 
%Based on this result, we will used the fixed value $\alpha = 0.8$ in our remaining experiments.

\subsection{Comparison between models}
In this second experiment, we will compare the performance of two proposed MILP models and the heuristic implementation in terms of solving time and optimal gap. Here,  we test each model on several network instances where the number of SNs and there are 20 network instances randomly generated for each network size. In this experiment, we also consider two different energy settings. In particular, for each network size, there are only 10 instances that take the default energy setting as presented above. With the 10 remaining network instances, each SN will take a random energy value from a more narrow interval [10J, 12J]. This second energy setting reflects the situation in a later round when the SN energies are gradually balanced through multiple previous rounds.
%for each network instance, namely the setting A where energy of a SN is randomly selected from the interval $[20, 40]$ and the setting B where the SN energies is a random value in a more narrow range $[10, 12]$. The first setting demonstrate the heterogeneous network scenario where the SNs can take different initial energy at the beginning of network operation, while the second setting reflect situation in later round when the energy of SNs are gradually balanced through multiple previous rounds.

We provide the comprehensive results in Table~\ref{tab:model_comparison}. In this table, the solving time and the best relative gap to the optimal solution, including the average and the worst case value, are respectively presented in the column "time" and "gap". Besides, the column "$P_{opt}$" shows the percentage of instances that can be solved optimally over 20 instances. Especially, a cell in column "gap" or "$P_{opt}$" is left blank in case that all test instances are solved optimally.

According to the results, 

\begin{table}[t]
	\caption{Comparison between models}
\begin{center}
\begin{tabular}{|c|c c r r r r|}
    \hline
    \multirow{2}*{N} & \multirow{2}*{Model} &
    \multirow{2}*{$P_{opt}$} &
    \multicolumn{2}{c}{Time (s)} & \multicolumn{2}{c|}{Gap (\%)} \\
     & & & \multicolumn{1}{c}{avg} & \multicolumn{1}{c}{max} & \multicolumn{1}{c}{avg} & \multicolumn{1}{c|}{max} \\ \hline
    \multirow{4}*{15} & 2-index & & 0.50 & 2.05 &  &  \\
     & 3-index & & 2.83 & 7.75 &  &  \\ 
     & Heuristic 50\% & & 0.13 & 0.20 &  &  \\ 
     & Heuristic 20\% & & 0.02 & 0.14 &  &  \\ \hline
     %\multirow{4}*{20} & 2-index & & 1.08 & 3.17 &  &  \\
     %& 3-index & 95\% & 14.95 & 59.99 &  &  \\ 
     %& Heuristic 50\% & & 0.41 & 0.80 &  &  \\ 
     %& Heuristic 20\% & & 0.10 & 0.34 &  &  \\ \hline
     \multirow{4}*{30} & 2-index & & 5.58 & 20.23 &  &  \\
     & 3-index & 45\% & 1026.02 & 1800.00 & 2.07 & 9.24 \\ 
     & Heuristic 50\% &  & 2.63 & 7.17 &  &  \\ 
     & Heuristic 20\% &  & 0.37 & 0.67 &  &  \\ \hline
     \multirow{4}*{50} & 2-index & 95\% & 256.66 & 1800.00 & 0.05 & 0.10 \\
     & 3-index & 0\% & 1800.00 & 1800.00 & & \\ 
     & Heuristic 50\% & & 38.62 & 122.77 &  &  \\ 
     & Heuristic 20\% & 95\% & 5.82 & 12.17 & 0.04 & 0.11 \\ \hline
     \multirow{4}*{100} & 2-index & & & & & \\
     & 3-index & & & & & \\ 
     & Heuristic 50\% & &  &  &  &  \\ 
     & Heuristic 20\% & & 126.61 & 360.09 &  &  \\ \hline
\end{tabular}
\end{center}
\label{tab:model_comparison}
\end{table}
%\vspace*{-15pt}

\subsection{Performance of the heuristic model}
To illustrate the effectiveness of our heuristic MILP approach, we compare it with the Convex-Hull-based protocol (CHP) in [10]. This protocol firstly divides the network into multiple sectors with the same percentage of energy decrease, and each mobile collector (UAV) is associated with a unique sector. A UAV trajectory is constructed inside its sector at initial by a geometrical method. Then a simulated-annealing adjustment is adopted to improve the trajectories in terms of total network energy consumption.

\begin{figure}[t!]
	\centering
	\setlength{\belowcaptionskip}{-10pt}
	\includegraphics[width=0.5\textwidth]{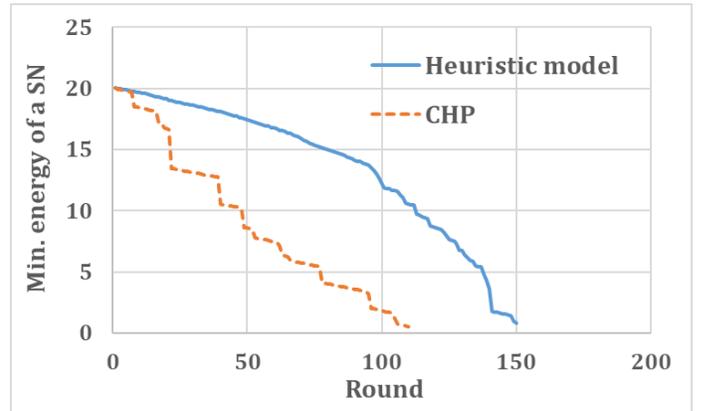}
	\caption{Network lifetime in multi UAV scenario}
	\label{lifetime-multi}
\end{figure}

Fig.~\ref{lifetime-multi} presents the minimum residual energy after each round when applying our heuristic model and the CHP in a network with 100 SNs. As can be seen, our algorithm has a lower rate of energy reduction. As a result, the network lifetime when applying the heuristic model is 150 rounds, higher than 36\% compare to 110 rounds as in the CHP. 

\section{Conclusion}
\label{sec:conclusion}

\section*{Acknowledgment}

This research is funded by Vingroup Joint Stock Company and supported by the Domestic Master/PhD Scholarship Programme of Vingroup Innovation Foundation (VINIF), Vingroup Big Data Institute (VINBIGDATA); funded by Vietnam National Foundation for Science and Technology Development (NAFOSTED) under grant number 102.02-2017.316.

\bibliographystyle{unsrt}
\bibliography{myref}

\end{document}